# Improving Transparency in IDIQ Contracts:

# A Comparison of Common Procurement Issues Affecting Economies Across the Atlantic and Suggested Solutions

Sareesh Rawat

The George Washington University Law School

# Table of Contents

# 1 ABSTRACT

Public procurement in Europe and United States has come a long way from simple traditional bid contracts under the English common law to complicated umbrella contracts. The expansion in scope and size of public procurement across the Atlantic has increased the need for responsibility and accountability of democratic governments. This has in turn lead to an increased demand for more transparent procurement systems. Umbrella contracts by their very nature are less transparent but serve as major tools for procurement in both Europe and United States. This paper brings up some common issues and challenges of reduced transparency under umbrella contracts and some ways of solving these issues. The paper begins by providing a brief history of IDIQ and Framework contracts and the growing demands of increased transparency under both systems. It then seeks to show why transparency must be balanced with efficiency in these contracts by generally discussing the importance of transparency in procurement systems. Finally, after reviewing certain ongoing challenges under the two systems, the paper concludes by discussing and providing potential solutions.

## 2 INTRODUCTION

A complex matrix of public procurement rules and regulations has developed in both Europe and United States because of several upheavals of corruption, nepotism and collusion.[1] The principle goals of this matrix include integrity, accountability, competition and transparency.[2] These goals must be constantly balanced with the need of speed and efficiency in procurement to ensure that government agencies are not overly burdened with administrative costs and contractors don't shy away from government business.[3] The development of Indefinite Delivery Indefinite Quantity (IDIQ) contracts and Framework Agreements from supplier lists in United States and Europe was a result of this balancing act.[4]

An increased level of spending and transactions, combined with a lack of experienced and adequately trained acquisition professionals has contributed to an increased use of IDIQ contracts. With an increased use of IDIQ contracts in the United States and Framework Agreements in the European Union, there has also been an increase in concerns of overuse and often misuse of these contracts.[5] While these contracts provide the government with greater flexibility and a more streamlined procurement process, the exclusivity and faster administration of these contracts leads to reduced transparency.[6]

---

[1] Christopher R. Yukins, *Are IDIQs Inefficient? Sharing Lessons with European Framework Contracting*, 37 Pub. Cont. L.J. 545 (2008).
[2] Wendy Ginsberg et al., *Government Transparency and Secrecy: An Examination of Meaning and Its Use in the Executive Branch*, 27-29 (Congressional Research Service, 2012).
[3] Yukins, *supra* note 1.
[4] *Id*.
[5] *Id*.

"Transparency goes beyond making more information available. Relevant and accurate information needs to be and accessible timely, especially to those who need it and can make use of it for policy or operations."[7] A comparison of IDIQs and Framework Agreements is possible due to their commonalities. Both aggregate demand for goods and services to be delivered at different moments in time and have a two stage procurement process. While IDIQs and frameworks also carry risks to competition and efficiency due to their inherent structure[8], they are often most criticized for reduced transparency. This is because transparency is a critical part of the procurement process as it promotes involvement of the public. It helps in equal treatment of contractors, as it ensures that the contracting authorities act in accordance with the law and do not offer preferential treatment to contractors. Transparency also assures compliance of other procurement principles. Transparency increases accountability and limits the scope for corruption and other fraudulent practices. It assures open competition and limits protectionism.[9]

---

[6] *Id.*
[7] Richard Golding, *Making the UN System more transparent and Accountable,* Future United Nations Development System (2014).
[8] *Id.*
[9] Yukins, *supra* note 1.

## 3 HISTORY OF TRANSPARENCY IN IDIQ AND FRAMEWORK CONTRACTS

The roots of American public procurement lie in England.[10] Initially government procurement was done primarily for the armed forces.[11] Government procurement gradually expanded to include goods, services and supplies.[12] Public procurement in both US and Europe has passed through different stages of development.[13] With increase in public procurement, the importance of ensuring equal and fair opportunity among competitors increased.[14] The needs for accountability and avoiding unnecessary red tape had to be balanced.[15] Moreover, with expansion of public procurement, the government agencies found the procurement process to be tightly regulated, time consuming and restrictive. So the public procurement in Europe and US slowly moved to Framework Agreements and IDIQs to overcome these impediments.[16]

Formation of the European Union in mid-1990's meant that integration of various members states and their economies became the primary European objective. Integration of public procurement was seen as one of the first steps to achieve this objective and framework agreements offered the desired flexibility for this purpose. Around the same time in the United States there was an increase in use of IDIQ contracts as a preferred method of procurement.[17]

---

[10] Sandy Keeney, *The Foundations of Government Contracting* (Citing James F. Nagel, *A History of Government Contracting*, George Washington University (1999)).
[11] *Id.*
[12] *Id.*
[13] Yukins, *supra* note 1.
[14] *Id.*
[15] Keeney, *supra* note 10.
[16] Yukins, *supra* note 1.

---

[17] *Id.*

# 4 TRANSPARENCY IN PROCUREMENT SYSTEMS

Transparency is identified as one of nine objectives frequently associated with government procurement.[18] and its suggested that system transparency is one of the three pillars (along with competition and integrity) that underlies the federal government's procurement system.[19] This section aims to show the value, costs and importance of transparency in procurement systems.

## 4.1 VALUE

Transparency is the paradigm without which neither fair competition nor accountability can be ensured. More opaqueness in a system increases the danger of less competition, and increases chances of collusion among competitors and far less accountability.[20] Transparency effects change through dissemination of information. More information provides chances for more constructive suggestions, criticisms and protests that may help to bring about change in the procurement process. By increasing participation, transparency works against discrimination. Transparency ensures improved integrity, accountability, equal opportunity, more compliance to rules and regulations, and more value for money. By exposing various corrupt practices, it improves the faith of people in the government procurement system. However, any efforts to improve transparency by expanding existing laws, or principles and exceptions to prevent collusion should be publicly stated. Arrowsmith gives examples from EU procurement system about the ways in which transparency rules may obstruct other procurement objectives by increasing costs.[21]

---

[18] Wendy Ginsberg, Maeve P. Carey, et al., Government Transparency and Secrecy: An Examination of Meaning and Its Use in the Executive Branch, Congressional Research Service, (November 14, 2012) (citing Prof. Schooner).
[19] *Id.*
[20] *Id.*

## 4.2 NEED

Where there is less transparency, there is probability that the government did not receive best value procurement.[22] Many vendors who have invested a lot in their marketplace relationships and the customer agencies who prefer less scrutiny, and quick, easy and cost effective services, prefer the opaqueness of these task orders.[23]

## 4.3 IMPORTANCE IN SPECIAL PROCUREMENT SCENARIOS

Government procurement is done for various types of requirements. Most procurements are standard but sometimes the government has to make emergency procurements. There is need for transparency at both information and process levels in all these types of procurements. Even when the government has to deal with such disasters like Hurricane Katrina, there is need for disclosing detailed information, "including requests for proposals, contract data, and contracting officers' decisions and justifications."[24] In the wake of Katrina disaster, emergency funding highlighted the special need for transparency. Prof. Yukins gives credit to the ever-vigilant media and high political pressure that made agencies their task orders more transparent.[25] Katrina disaster showed that transparency in providing details of task orders under IDIQs is possible without compromising or affecting the procurement process.[26] Disaster like "Katrina reminds us that transparent task-order contracting not only is possible, but it makes the procurement system stronger and more accountable."[27]

---

[21] S. Arrowsmith, *The EC Procurement Directives, National Procurement Policies and Better Governance: The Case for a New Approach* 27 Euro. L. Rev. 3 (2002).
[22] Steven L. Schooner, Christopher R. Yukins, "Emerging Policy and Practice Issues (2005)", GW Scholarly Commons.
[23] *Id*.
[24] *Federal Contracting: Lessons Learned from hurricane Katrina*, Project On Government Oversight (2006).

---

[25] Christopher R. Yukins, *Hurricane Katrina Brings Transparency to Task Order Contracting*, (Contract Services Ass'n, 2006).
[26] *Id.*
[27] Steven L. Schooner and Christopher R. Yukins, *Emerging Policy and Practice Issues* (2005).

## 5  COMMON ISSUES OF TRANSPARENCY IN FRAMEWORK AGREEMENTS IN EUROPE

Framework agreements can be broadly categorized into single contractor and multiple contractor agreements. While issues of transparency certainly exist in single contractor framework agreement, the mini-competition stage of the multiple contractor framework agreements poses more issues.[28] Article 33 of the EU procurement directive of 2014 requires contracting authorities to "consult in writing, the economic operators capable of performing the contract."[29] This broad requirement has led to ambiguities as to what stage of the framework agreement should the contractor's capability to perform the contract be assessed?[30] If transparency were the primary interest, contractor capability would be assessed at the time of the framework agreement.[31] However, it is often more practical to determine contractor capability at the time of the mini-competition as the specific requirements of the task-order may not be available at the time of the drafting of the framework agreement.[32] This language of Article 33 has also been criticized for being open to multiple interpretations regarding which contractors the contracting authorities must consult.[33] There is an argument that all contractors that are part of the agreement at the framework level should automatically be assumed as capable of performing the contract, and therefore must all be consulted.[34] However, the language has also been interpreted to give the contracting authorities discretion to choose between the various contractors that are party to the framework agreement, and only consult those contractors that they deem capable of performing the specific task order.[35]

---

[28] Marta Andrecka, *Framework Agreements, EU Procurement Law and the Practice*, 2 Procurement L.J., 2015.
[29] Procurement Directive 2014/24 art. 33(5) (EU).
[30] Marta Andrecka, *Framework Agreements: Transparency in the Call-off Award Process*, 10 Eur. Procurement & Pub. Pvt. Partnership L. Rev. 2015.
[31] *Id*.
[32] *Id*.
[33] *Id*.

Article 50 of the EU procurement directive poses a significant transparency issue in framework agreements concluded in accordance with Article 33 of the directive. Under this article, contracting authorities have no obligation to notify the participants of the framework agreements of the results of the mini competitions under the agreement.[36] This lack of obligation to notify unsuccessful contractors within a framework agreement, when coupled with the ambiguous language of Article 33 may pose significant risks to transparency.[37]

---

[34] *Id.*
[35] *Id.*
[36] Procurement Directive 2014/24 art. 50(2) (EU).
[37] Andrecka, *supra* note 29.

# 6 COMMON ISSUES OF TRANSPARENCY IN IDIQ CONTRACTS IN THE UNITED STATES & RECOVERY ACT SOLUTIONS

It was not until the 1966 enactment of the Freedom of Information Act (FOIA; 5 U.S.C. § 552) that individuals, corporations, and other entities were given "presumptive access to unpublished, existing and identifiable records of the agencies of the Federal executive branch without having to demonstrate a need or reason for such request. [38] The American Recovery and Reinvestment Act of 2009 was introduced to stimulate the American economy after the recession of 2008. The Recovery Act had transparency provisions that allowed the public to track the rate of spending and recovery due to the Act.[39]

The Office of Management and Budget outlined several accountability goals under the Recovery Act of 2009. These goals impact transparency in procurement under IDIQ contracts as they require that all funds are disbursed and reported clearly, accurately and in a timely manner.[40] FAR Subpart 5.7 prescribes the posting requirements under the American Recovery and Reinvestment Act of 2009.[41] The publishing requirements of the Recovery act are applicable to task-order contracts, awarded in the second phase of IDIQ contracts under FAR 16.5.[42] The Recovery Act took significant steps in improving transparency by expanding posting requirements for pre-solicitation and post-award phases for actions funded in whole or part by the

---

[38] Wendy R. Ginsberg, *The Obama Administration's Open Government Initiative: Issues for Congress"* (*Congressional Research Service,* 2011).
[39] American Recovery and Reinvestment Act of 2009, H.R. 1, 111th Cong. (2009).
[40] OMB Memorandum "Initial Implementing Guidance for the American Recovery and Reinvestment Act of 2009" (M-09-10), Peter Orszag, February 18, 2009.
[41] FAR 5.7.
[42] FAR 16.5.

Act. For pre-award publicizing, in addition to the publication procedures in FAR 5.201, the Recovery Act requires contracting officers to post informational notices for public use for all actions exceeding $25,000. Similarly, under post-award requirements, the government must publicize all actions exceeding $500,000 funded in whole or in-part by the Recovery Act. This requirement is a critical expansion of transparency in procurements done using IDIQ contracts and Federal Supply Schedules, as it supersedes the exceptions to notice requirements in FAR 5.301(b)(2)-(7), which included task orders under multiple-award contracts, as long as they met the fair opportunity standards outlined in FAR 16.5.5(b)(2). The post-award notice requirements also apply to modifications of existing contracts or task-orders.

In addition to the pre-and post-award notice requirements, in cases where a contract or a task order is not competitively awarded or is not fixed price, the Recovery Act requires contracting officers to post a rationale supporting their decision which is available to the public.[43] Furthermore, a minimum dollar amount threshold does not bind the supporting rationale requirement.

---

[43] *Id*.

# 7 MORE SOLUTIONS TO IMPROVE TRANSPARENCY IN IDIQ AND FRAMEWORK CONTRACTS

## 7.1 WORLD BANK'S NEW PROCUREMENT FRAMEWORK

An independent evaluation conducted by OECD in 2010 found that "opaque tendering in large lots" was a primary reason behind the challenges faced by micro, small, and medium enterprises (MSMEs) in competing for tenders in developing countries.[44] In 2015, World Bank approved a new public procurement framework for all the projects financed by it.[45] This new procurement policy "impacts a portfolio of about USD 42 billion in over 1800 projects in 172 countries."[46] While best value for money remains the core procurement principle in the new framework, World Bank has started preferring bids with overall best value for money over lowest cost bids.[47] The UNCITRAL Model Law of Procurement also recognizes the danger of compromising on value for money, where Framework Agreements are used beyond their scope for administrative convenience.[48] In order to increase transparency, World Bank has also introduced the concept of standstill period, which requires a pause period between selection of the winning contractor and the actual award of the contract.[49] This allows other bidders the opportunity to voice any concerns before the contract is actually legally formed and enforced.[50] UNCITRAL Model Law of Procurement and EU Procurement Directives also employ similar standstill measures to increase accountability through increased transparency.[51]

---

[44] Jeroen Kwakkenbos, *Assessing the World Bank's New Procurement Policy* (European Network of Debt and Development, 2015).
[45] *Id.*
[46] *Id.*
[47] WORLD BANK PROCUREMENT FRAMEWORK, § III C (THE WORLD BANK GROUP, 2016) (explaining the seven core principles of World Bank Procurement, namely, 1. Value for money, 2. Economy, 3. Integrity, 4. Fitness for Purpose, 5. Efficiency, 6. Transparency, and 7. Fairness).
[48] Gian Luigi Albano & Caroline Nicholas, at §7.4
[49] Kwakkenbos, *supra* note 26.

## 7.2 UNCITRAL MODEL LAW ON PUBLIC PROCUREMENT

Article 4 of the UNCITRAL Model Law on Public Procurement provides guidance on government procurement, recognizing transparency as one of its primary goals.[52] UNCITRAL Model law recommends that agencies only use framework agreements in cases where the subject matter of the procurement arises on an indefinite basis during a specific period of time, or if there is an urgent need of the agency to procure.[53] Similar to the supporting rationale requirements of the Recovery Act, UNCITRAL also requires that contracting authorities record their reasons and circumstances that justify the use of a framework agreement.[54] Such limitations on use of framework agreements can be a potential solution to overuse.[55] However, it is critical that any such limitations, as well as rights and obligations of the contracting parties be clearly articulated either as formal rules or internal rules within the agency as applicable, to support transparency and accountability.[56] If done properly, these guidance may arguably serve an additional purpose by being effective accountability measures.[57]

---

[50] *Id*.
[51] Gian Luigi Albano & Caroline Nicholas, at §7.4.

[52] UNCITRAL MODEL LAW ON PUBLIC PROCUREMENT, art. 32 G.A. Res. 2205 (XXI), Annex, U. N. GAOR, 66th Sess., Supp. No. 17, U.N.Doc. A/66/17 (U.N. COMM'N ON INT'L TRADE LAW, 2011). [hereinafter UNCITRAL Model Law].
[53] *Id*.
[54] UNCITRAL Model Law *supra* note 22, art. 32(2).
[55] Caroline Nicholas, 'A Critical Evaluation of the Revised UNCITRAL Model Law Provisions on Regulating Framework Agreements', Public Procurement Law Review2 (2012): 19–46

Furthermore, the UNCITRAL Model law contains dedicated procedures for the mini-competition stage, unlike the EU and US systems, that "impose transparency requirements, so that only awardees can take part in the call-off competition."[58] It emphasizes the importance of proper procurement planning before each stage of Framework Agreements to ensure adequate selection, operation and conclusion.[59] It suggests that procurement agencies adhere to certain safeguards to promote transparency, such as, "advance notice of the procurement, publishing notices of procurement contract awards and keeping comprehensive records of procurement proceedings."[60] UNCITRAL also encourages the use of e- procurement for increasing transparency.[61]

## 7.3 E-PROCUREMENT & DYNAMIC PURCHASING SYSTEMS

One use of e-procurement that can particularly benefit EU's current system, involves the increased use of Dynamic Purchasing Systems over framework agreements under appropriate conditions. This is due to the fact that the EU has generally smaller value of contracts when compared to the contracts in the United States. Dynamic Purchasing Systems are relatively infrequently used in EU procurement but are similar to Multiple Award Schedules made popular by the General Services Administration (GSA).[62]

---

[56] Gian Luigi Albano & Caroline Nicholas, The design of framework agreements, in The Law and Economics of Framework Agreements: Designing Flexible Solutions for Public Procurement § 12.4.2 (2016).
[57] *Id*.
[58] *Id*. at § 7.4.
[59] Guidance on Procurement regulations to be promulgated in accordance with article 4 of the UNCITRAL Model Law on Public Procurement http://www.uncitral.org/pdf/english/texts/procurem/ml-procurement-2011/Guidance-on-procurement-regulations-e.pdf
[60] *Id*.
[61] *Id*.

Similar to MAS contracts where the master agreements are "always open" to new competitors,[63] these systems allow suppliers to join the contract at any time during the period of performance.[64] This potentially solves one of the biggest issues of competition, and as a result transparency, under framework agreements. Professor Arrowsmith describes dynamic purchasing systems as contracts that "authorizes entities to establish, using electronic means, a list of suppliers interested in supplying certain standard supplies or services."[65] To join the Dynamic Purchasing System, the contractor must first register.[66] This is done by submitting a mandatory and compliant bid for the relevant product or service.[67] Any firm that submits a compliant proposal must be admitted to the system.[68] However, there are trade-offs in costs and efficiency of procurement. "When the procuring entity wishes to place an order under the system, it cannot simply select a tender from the system, but must place a new simplified notice of the dynamic purchasing system in the *Official Journal*, then allow new suppliers to register, and then seek tenders for the particular order from all the registered suppliers.[69]

---

[62] Christopher R. Yukins, *Are IDIQs Inefficient? Sharing Lessons with European Framework Contracting*, 37 Pub. Cont. L.J. 545, 2008
[63] *Id.*
[64] Gian Luigi Albano, Antonio Ballarin and Marco Sparro, *Framework Agreements and Repeated Purchases: The Basic Economics and a Case Study on the Acquisition of IT Services*, (International Public Procurement Conference, 2009).
[65] Yukins, *supra* note 1 (citing Professor Arrowsmith's description of Dynamic Purchasing Systems).
[66] *Id.*
[67] *Id.*
[68] *Id.*

---

[69] *Id*.

# 8 CONCLUSION

Both systems can benefit from greater communication and sharing lessons learned, even if the lessons are small. For instance, framework agreements are generally for no longer than four years. The American IDIQs could benefit from a similar arrangement as multiple-award IDIQ contracts are often ten years or longer. This could lead to the contract terms being obsolete and could contribute to either under-utilization or insufficiency of the IDIQ ceiling amount. An even shorter period for 1-2 years can be considered at the framework agreement level to soften the crushing blow suffered by small and medium sized business from not getting on the IDIQ contract. Task orders can still be 5-10 years as usual. Certain other measures at different stages of procurement, such as, transparency at the time of calling for bids, a public notice requirement, government briefings to competing contractors, continuous monitoring of umbrella contracts throughout the period of their operations can benefit both systems.

Communication and discussion of ideas especially those of eminent scholars in the field is critical for increasing transparency and generally improving the two systems. For instance, Daniel I. Gordon advocates the right to protest bids to help to improve transparency.[70] Another example is Prof. Yukins' Principal-Agent model, which has lent a new perspective on transparency in public procurement by utilizing principal agent theory. According to the model, "theoretically the principal can ensure better outcomes from an agent if the agent is afforded more complete information on the agent's own performance."[71] Implementation of this perspective would provide an additional efficiency based incentive for increased transparency in IDIQ contracting, provided the dissemination of information is properly balanced with confidentiality. Essentially, with increased transparency, "the "principal" guiding procurement may no longer be the head of an agency, or even Congress—it may, in time, be the end users (the veteran or the rural community) who are considered the "principals," with a first say in how a procurement should be shaped."[72] Issues as broad as transparency in umbrella contracts are rarely solved overnight and require a free flow of ideas across systems.

---

[70] Daniel I. Gordon, *Protests: The Costs Are Real, But the Benefits Outweigh Them*, 42 PUB.CONT. L.J 489, 492 (2013).

---

[71] Christopher R. Yukins, *A Versatile Prism: Assessing Procurement Law Through the Principle- Agent Model*, 40 Pub. Cont. L. J. 1 (2010)
[72] *Id*.